\documentclass[%
 reprint,
 amsmath,amssymb,
 aps,
pre,
]{revtex4-2}

\usepackage{graphicx}
\usepackage{dcolumn}
\usepackage{xcolor}
\usepackage{bm}

\begin{document}

\preprint{APS/123-QED}

\title{Conformation and topology of cyclical star polymers}

\author{Davide Breoni}
 \email{davide.breoni@unitn.it}
\affiliation{Soft matter and Biophysics institute, Università di Trento, Via Sommarive 14, 38123 Trento, Italy }
\author{Emanuele Locatelli}%
\affiliation{Department of Physics and Astronomy, University of Padova, Via Marzolo 8, I-35131 Padova, Italy}
\affiliation{INFN, Sezione di Padova, Via Marzolo 8, I-35131 Padova, Italy}
\author{Luca Tubiana}
\affiliation{Soft matter and Biophysics institute, Università di Trento, Via Sommarive 14, 38123 Trento, Italy }

\begin{abstract}
We study the conformation and topological properties of cyclical star polymers with $f$ ring arms, each made of $n$ beads. We find that the conformational properties of unlinked cyclical star polymers are compatible to those of linear star polymers with $2f$ arms made of $n/2$ beads each. This compatibility vanishes when the topology of the star, measured as the degree of linking between arms, changes. In fact, when links are allowed
we notice that the gyration radius decreases as a function of the absolute linking number $\vert Lk \vert$ of the arms, regardless of the protocol that is employed to introduce said links. 
Furthermore, the internal structure of the macromolecules, as highlighted by the radial density function, 
changes qualitatively
for large values of $\vert Lk \vert$. 
\end{abstract}

\maketitle

\section{\label{sec:intro}Introduction.}
Polymeric materials are ubiquitous in materials science and biology. Their success is due to their versatility, as their properties, even in diluted conditions, depend on their chemical composition (homo-, multiblock- or hetero-polymers), physical architecture and topology~\cite{bates201750th,liang201750th,polymeropoulos201750th,blasco201750th,chen201750th,tubiana_topology_2024}. 
In particular, star polymers~\cite{grest1996star}, characterized by a certain number of arms grafted to a central core, emerged as a particularly interesting model for soft systems\cite{likos2001effective}. First, their synthesis is well controlled~\cite{zhou1993synthesis,polymeropoulos201750th}, posing them as a reliable experimental system~\cite{vlassopoulos2004colloidal}. Second, from the theoretical perspective, they provide a link between polymeric and colloidal physics\cite{vlassopoulos2014tunable}: stars with very few arms  resemble linear chains while, in the high $f$ limit, they can be described as sterically stabilized colloids. Third, their very symmetric nature makes them quite suitable for the application of coarse-graining techniques~\cite{likos2001effective,jusufi2001effective,mayer2007coarse,d2015coarse,pierleoni2007soft,marzi2012coarse,locatelli2016multiblob}, that yield effective interaction retaining predictive power beyond the glass transition~\cite{foffi2003structural}. Further, the effective interaction has a characteristic ``ultrasoft'' nature~\cite{likos_star_1998}, that leads to rich phase diagrams for pure systems~\cite{watzlawek1999phase,foffi2003structural,menichetti2017thermodynamics} as well as for mixtures~\cite{dzubiella2001phase,camargo2010unusual,parisi2021effect,parisi2023gelation,truzzolillo2013glassy,marzi2015depletion}. Finally, the star polymer architecture also offer the possibility for easy external control, as the central core can be made with a magnetic material~\cite{roma2021theoretical}, decoupling 
their transport from the material properties.\\A completely different, yet also archetypal, way of macromolecular organization is a polymer ring. Albeit conceptually much simpler than stars, as they can be obtained from simple linear chains by adding just one bond, rings have fascinated and puzzled scientists for decades\cite{tubiana_topology_2024, wasserman_chemical_1962}. Indeed, their chemical synthesis is far from trivial; reliable methods have been developed only recently~\cite{haque2020synthesis,chen2022cyclic}. From a more physical perspective, closing a linear chain with one bond effectively eliminates the chain ends: with their disappearance, a striking amount of properties changes with respect to the linear case, both at the level of single molecules~\cite{mai2016single,hsiao2016ring,liebetreu2020hydrodynamic,tubiana_topology_2024} and in macroscopic rheology\cite{kapnistos:NatMat:2008,Tsalikis2017,Rosa_Everaers_PRL_2014,halverson:jcp1:2011,Michieletto_Turner_topoglass_PNAS16,tubiana_topology_2024}. Indeed, what is introduced is a \emph{topological} constraint that, for example, induces an effective swelling, even in the absence of excluded volume~\cite{Grosberg:2000:PRL}, an effective stronger repulsion between different polymers~\cite{narros2010influence} or between a polymer and a wall~\cite{chubak2018ring}. Further, one can use the topological constraint to \emph{link} rings together\cite{orlandini2021topological}, building topologically complex structures as catenanes~\cite{dehaghani2023topological,tubiana2022circular,farimani2024effects}, or topological materials as Olympic gels\cite{vilgis1997elasticity,fischer2015formation,krajina2018active,hart2021material} and the KDNA~\cite{chen1995topology,he2023single,tubiana_topology_2024}.\\On the other hand, the interesting interfacial properties of cyclical polymer, grafted on solid surfaces to form a brush, have emerged only recently ~\cite{morgese2016topological,divandari2017,romio_topological_2020}. For example, cyclical macromolecules show enhanced
steric stabilization, impacting nonspecific adsorption as well as lubrication properties~\cite{morgese2016topological,divandari2017,morgese2018cyclic}. For some applications, spherical cyclical brushes, i.e. cyclical star polymers, have been proposed and synthesized~\cite{morgese2017next}. While there are studies on planar ring brushes~\cite{Reith_2011,Jehser2019}, so far nobody investigated the conformational and the scaling properties of cyclical stars from a theoretical perspective. Further, the effect of additional topological constraints, in the form of links, on the grafted rings has never been considered.\\In this work, we consider cyclical star polymers in good solvent conditions and study their conformational and scaling properties using numerical simulations. Employing the Kremer-Grest bead-spring model\cite{grest_molecular_1986}, we compare how the average size of cyclic star polymers compares to their linear counterpart upon suitably varying the number and the length of the grafted chains. Further, we consider two ways to effectively introduce topological entanglements (links) in the system: i) by temporarily turning off the self-avoidance among arms and, in addition, by imposing a radial confinement to the system; ii) by considering ``doubly grafted'' linear chains, i.e. chains that are grafted on the central core at both ends that, via a simple geometrical rule, can be randomly arranged, yielding a certain degree of linking. In both cases, we investigate how the degree of linking affects the overall shape of the macromolecule. The paper is organized as follows: in Section~\ref{sec:methods} we discuss the model and simulation details (Sec.~\ref{ssec:simulations}), the generation of the initial configurations (Sec.~\ref{ssec:init}), the protocol used to set the spherical confinement, used to induce linking (Sec.~\ref{ssec:compr}), the definition of gyration radius and asphericity (Sec.~\ref{ssec:asph}), and, finally, the calculation of the linking number in the different cases (Sec.~\ref{ssec:linkn}). Then, in Section~\ref{sec:results} we discuss our results: in Sec.~\ref{sec:unlinked}, we compare the scaling properties of star polymers with either linear or cyclical arms, considering the metric properties  (\ref{ssec:gyration}) and the radial density (\ref{ssec:radial}). Further, in Sec.~\ref{sec:linked} we will introduce and discuss two different ways to effectively introduce links in the system and discuss the effects of linking on the metric properties (gyration radius and asphericity, Sec.~\ref{ssec:compression}, \ref{ssec:anchoring}), as well as on the scaling properties (Sec.~\ref{ssec:lkscaling}). Finally, we summarize and draw
our conclusions in Section~\ref{sec:outro}.
\section{\label{sec:methods}Methods.}
\subsection{\label{ssec:simulations} Numerical model and simulation details.}
We carry out coarse-grained numerical simulations with LAMMPS, where we simulate in 3D star polymers with a number of branches -- called functionality --  $f$ ranging from 20 to 300 for linear star polymers and from 10 to 150 for cyclical ones, with each arm having a number of beads $n$ between 10 and 500 for linear star polymer and between 20 and 1000 for cyclical ones. The size of the core diameter is fixed at $\sigma_c=8\sigma$, where $\sigma$ is the diameter of all non-core beads, and its mass is correspondingly set to $m_c=\left(\frac{\sigma_c}{\sigma}\right)^3m$, where $m$ is the mass of a non-core bead. All beads repel each other with a Weeks-Chandlers-Anderson (WCA) potential $U_r^{ab}(r)$, where $a$ and $b$ determine the type of the interacting particles (either core or non-core), shifted of an amount $\Delta^{ab}$ to take into account different bead sizes :
\begin{equation}
\label{Ur}
\resizebox{1\hsize}{!}{%

$U_r^{ab}(r)=
\begin{cases}
4\epsilon \left[\left(\frac{\sigma}{r-\Delta^{ab}}\right)^{12}-\left(\frac{\sigma}{r-\Delta^{ab}}\right)^6\right]+\epsilon, &r\leq r_{c}+\Delta^{ab},\\
0 ,&r> r_{c}+\Delta^{ab},
\end{cases} $

}
\end{equation} 
where $r$ is the distance between beads, $r_{c}=\sqrt[6]{2}\sigma$, $\epsilon$ is the energy scale of the system and the shift between two beads of diameter $\sigma_a$ and $\sigma_b$ is $\Delta^{ab}=(\sigma_a+\sigma_b)/2-\sigma$.
 The polymers are modeled within the Kremer-Grest framework, where consecutive beads are connected via finite-extensible non-linear elastic (FENE) springs $U_s^{ab}(r)$, again shifted to accomodate different bead sizes:
\begin{equation}
\label{Us}
U_s^{ab}(r)=
\begin{cases}
-\frac{KR_0^2}{2}\text{ln} \left[1-\left(\frac{r-\Delta^{ab}}{R_0}\right)^2\right],&r\leq R_0+\Delta^{ab},\\
\infty, &r>R_0+\Delta^{ab},
\end{cases}
\end{equation}
where $K=30\epsilon/\sigma^2$ is the stiffness of the spring and $R_0=1.5\sigma$ is the maximum bond length. In the case of the ``doubly grafted'' arms, the anchoring points are kept at a fixed position with respect to the central core; we thus treat the collection of the anchoring points plus the central bead as a rigid body. In all other cases, the same anchoring points are free to diffuse an rearrange on the surface of the core. For each bead $i$ with two neighbors (except the core), rigidity is introduced with the bending potential $U_{b,i}$:
\begin{equation}
\label{Ub}
U_{b,i}=
\kappa (1+\cos{\theta_i}),
\end{equation}
where $\theta_i$ is the angle formed by beads $i-1$, $i$ and $i+1$ and $\kappa=\epsilon$ is the bending energy. The system evolves with a Langevin thermostat, with temperature $T=\epsilon/k_B$ and damping coefficient $\gamma=\tau$, where $\tau=\sigma \sqrt{m/\epsilon}$ is the unit of time of the system. The time-step is $\text{d}t=0.01\tau$. For linear and unlinked cyclical stars we perform a single, very long simulation run of $10^6 \tau$, starting from a well equilibrated conformation. For linked stars, since we need to average over different realisations of the arms' topology, we perform, after reaching a steady state (see more details below), $M=$10 slightly shorter runs ($10^5 \tau$) for each different set of values of the parameters considered.   
\begin{figure}
\includegraphics[width=0.5\textwidth]{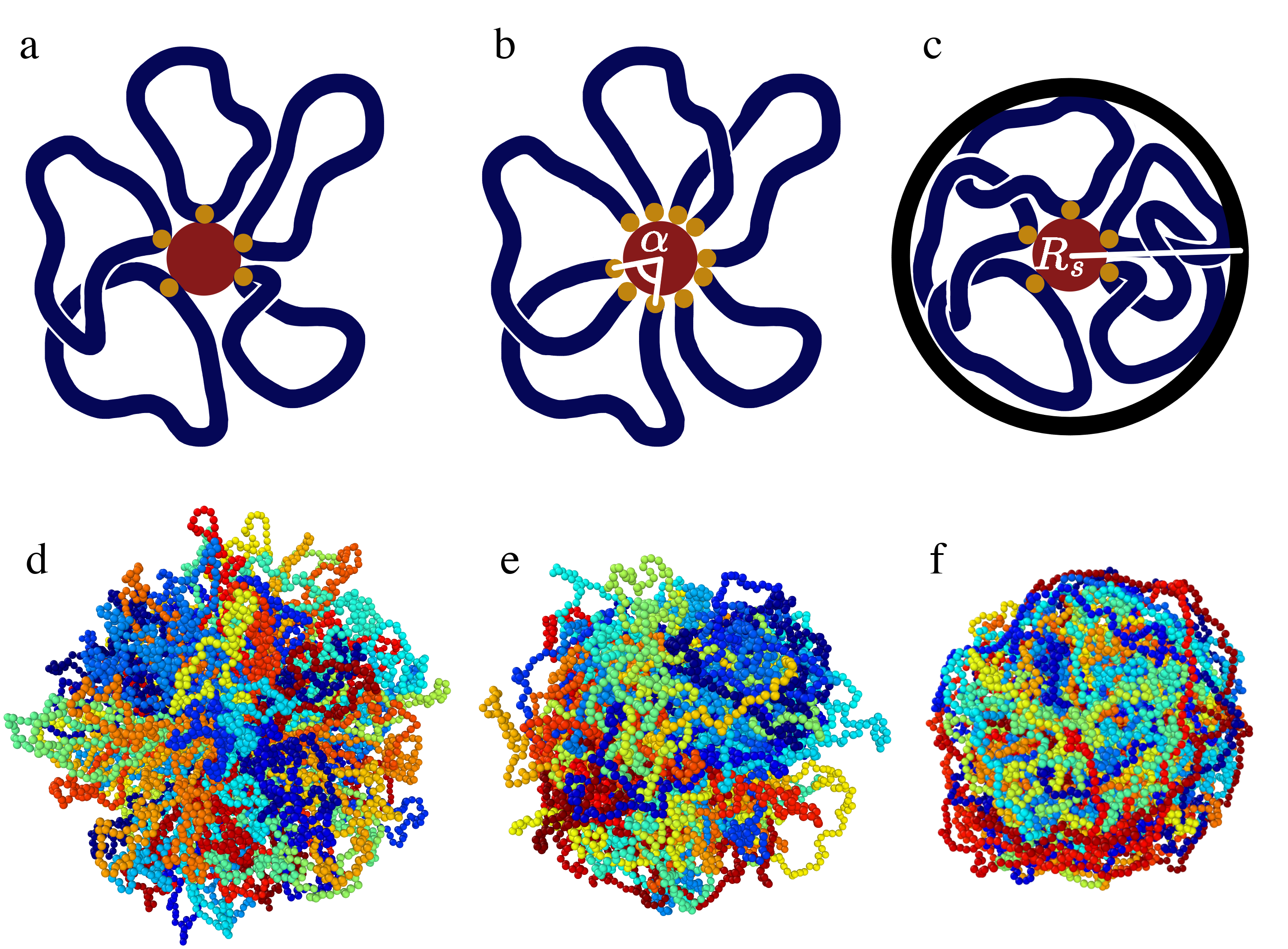}
\caption{\label{fig:daisy} a-c) Sketches of cyclical star polymers, made of a central core (in red), on which $f$ rings are attached (in blue) via beads (in orange). The stars are respectively a) unlinked, b) linked via doubly-grafted arms and c) linked via confined phantom arms. In panel b), $\alpha$ is the angle between anchoring points of the same arm; in panel c) $R_s$ is the confinement radius.   d-f) Typical snapshots of cyclical star polymer simulations, where each arm is shown in a different color. These example molecules are made of $f=100$ ring arms, with $n=100$ beads each. The polymers in the snapshots are respectively d) unlinked, e) doubly grafted and f) compressed stars.}
\end{figure}
\subsection{\label{ssec:init}Initial configurations.}
In order to cover the surface of the core in the most uniform way possible, and to avoid conflicts between beads, arms are initially anchored to the surface of the core in a spherical Fibonacci spiral (also known as {\it sunflower seed pattern}) \cite{gonzalez_measurement_2010}. In the case of linear stars and unlinked cyclical stars (i.e. stars where the arms are rings anchored to a single site of the core, see Fig.\ref{fig:daisy}a,d), we initially place the $f$ arms along the pattern; anyway, as mentioned above, they will diffuse on the surface of the core and rearrange in time. The arms are initially set as radially expanding lines from the core; one single chain for linear stars and two parallel chains connected at their extremities for cyclical stars. In the case of doubly grafted arms (see Fig.\ref{fig:daisy}b,e and Fig.\ref{fig:Lk_sketch}b), we first create the Fibonacci distribution of anchoring points on the core, then for each arm we randomly choose its two grafting sites such that the angle between them with respect to the center of the core is close to a set angle $\alpha$ . 
Once the two sites are chosen, we construct the arm by growing two chains radially outwards and, once they reached a certain length, which depends on their total length and on the angle $\alpha$, we connect them  with a spherical arc. 
As mentioned, in order to avoid changes in the obtained topology during the simulation, the anchoring points  are grafted to the core, so that the core and anchoring points form a rigid body. Thanks to the randomness in the procedure, we can sample a large variety of topologies; we performed ten realisations for each parameter set.
\subsection{\label{ssec:compr}Setting the confinement.}
One of the methods we employed to induce linking in cyclical stars involves the confinement of the star and the removal of self-avoidance among arms (see Fig.\ref{fig:daisy}c,f). The confinement protocol consists of three steps. First, a spherical hard boundary is set around the star, with a radius larger than the maximum possible extension of the star. As soon as the repulsion among non-neighbouring non-core monomers is removed, the spherical confinement slowly reduces its radius, until a prescribed radius $R_s$ is reached. In the second step, the star polymer is allowed to evolve within the spherical confinement until equilibrium is reached. The latter is verified through the time dependence of the gyration radius. Finally, the confinement is removed and repulsion among beads is slowly reintroduced by means of an initially soft potential, which repulsion grows stronger over time. For this we use the LAMMPS {\it soft} potential $U_{soft}(r)$: 
\begin{equation}
    U_{soft}(r)=s\left[1+\cos\left(\frac{\pi r}{r_c}\right)\right] \qquad \text{for } r<r_c,
\end{equation}
where we gradually increase the parameter $s$ from $s=\epsilon$ to $s=50\epsilon$ over $5 \cdot 10^4$ 
time steps, at which point $U_{soft}$ is switched to WCA.
Once this procedure terminates, the production run starts and we collect the data. This procedure is repeated ten times for each parameter set, in order to sample different realization of the linked star polymer topology.
\subsection{\label{ssec:asph}Metric properties.}
We consider here two observables, specifically the {\it gyration radius} $R_g$ and the {\it asphericity} $A$. The gyration radius $R_g$ is defined as:
\begin{equation}
\label{rg}
    R_g \equiv \left(\frac{1}{N-1}\left\langle\sum_{i=1}^{N-1} (\textbf{r}_i-\textbf{r}_{0})^2\right \rangle\right)^{1/2} - \frac{\sigma_c}{2}, 
\end{equation}
where $N=f\cdot n +1$  is the total number of beads $\textbf{r}_i$ is the position of $i$-th monomer, $\textbf{r}_0$ is the position of the core and $\langle\cdot\rangle$ is the ensemble average, computed over time along a single, long trajectory (see Sec.~\ref{sec:methods}) and, if necessary, over the ensemble of $M$ independent realizations.\\
The asphericity is computed employing a modified moment of inertia tensor $T_{\alpha\beta}$, which we define as
\begin{eqnarray}
\label{Tab}
    T_{\alpha\beta} &\equiv& \frac{1}{N-1}\left\langle\sum_{i=1}^{N-1} (r_{\alpha,i}-r_{\alpha,0})(r_{\beta,i}-r_{\beta,0})\right \rangle,
\end{eqnarray}
where $\alpha$ and $\beta$ refer to the Cartesian axes. This modified tensor is calculated in the reference system of the core bead $\textbf{r}_0$, and not the center of mass, as we are interested in deviations from a spherical polymer centered in the core.\\
From $T_{\alpha\beta}$ we extract the semiaxes of the ellipsoid enveloping the star polymer by taking the square root of the three eigenvalues of $T_{\alpha\beta}$. With these semiaxes ($a$, $b$ and $c$), we can calculate the asphericity $A$ \cite{millett_effect_2009}:
\begin{eqnarray}
\label{as}
    A\equiv \frac{(a-b)^2+(b-c)^2+(c-a)^2}{2(a+b+c)^2}.
\end{eqnarray}
This quantity measures how spherical the enveloping ellipsoid is: $A=0$ indicates a perfectly spherical shape, while a rod-like ellipsoid yields $A=1$.
\subsection{\label{ssec:linkn}Linking number.}
\begin{figure}
\includegraphics[width=0.5\textwidth]{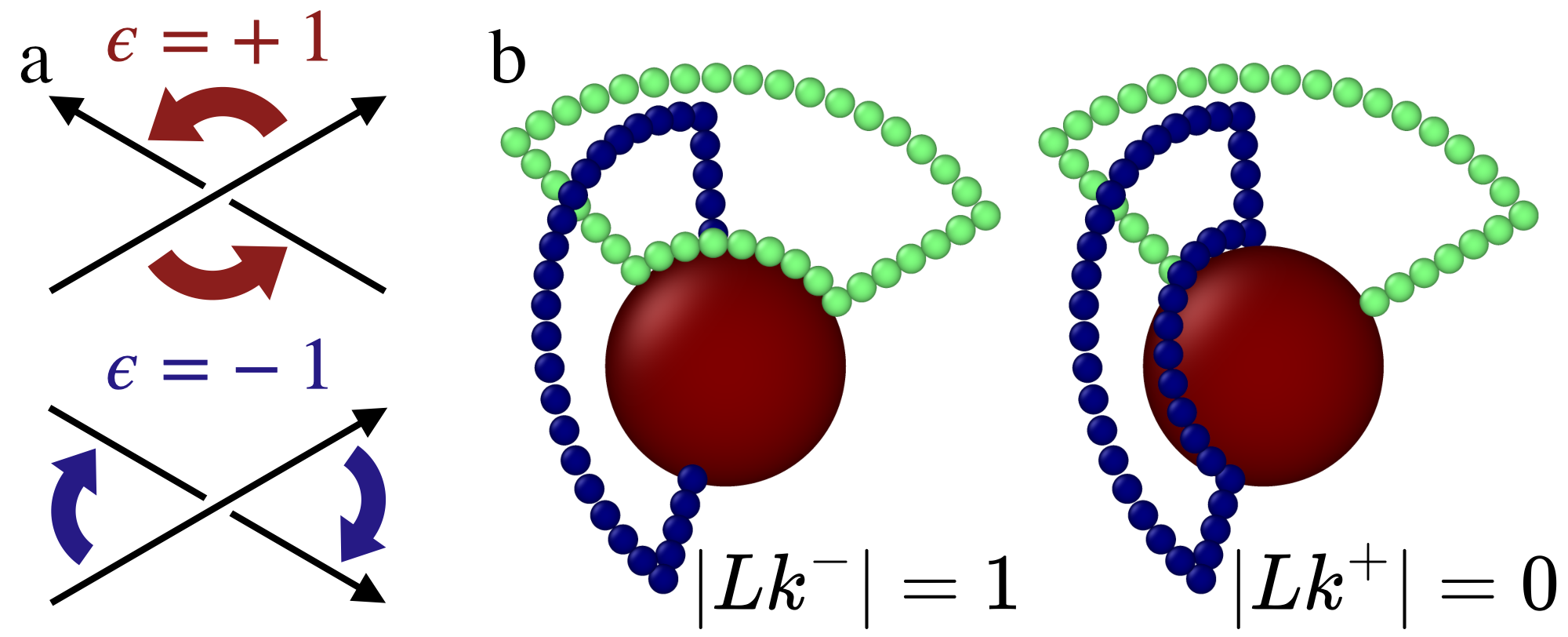}
\caption{\label{fig:Lk_sketch} a) Sign $\epsilon$ of a crossing point based on the verse of the two crossing strands. b) Closure procedure for two doubly-grafted arms: one arm is closed through the core center while the other is closed over the core surface, for both possible configurations. The values of the absolute linking number calculated for the two different closures (here shown as $\vert Lk^-\vert$ and $\vert Lk^+\vert$) are then averaged to obtain $\vert Lk\vert=(\vert Lk^-\vert+\vert Lk^+\vert)/2$.}
\end{figure}

Given two  closed oriented curves in 3D $C_1$ and $C_2$, their {\it linking number} $Lk(C_1,C_2)$ is defined as \cite{tubiana_topology_2024}:
\begin{equation}
    Lk(C_1,C_2)=\frac{1}{2} \sum_{p\in C(C_1\sqcap C_2)}\epsilon(p),
\end{equation}
where $C(C_1\sqcap C_2)$ is the set of all points in which the two curves cross each other under a certain 2D projection and $\epsilon(p)=\pm 1$ depending on if  the crossing point $p$ is positive or negative within the same projection. A crossing is positive if the overpassing strand can be superimposed to the underpassing one by rotating it counterclockwise. It is negative if it needs to be rotated clockwise (see Fig.\ref{fig:Lk_sketch}a). 
 In order to measure the linking number between the arms of the polymer, we first simplify the configuration by removing all topologically unnecessary nodes  through {\it rectification}; this is done with a modified version of the KymoKnot program \cite{tubiana_kymoknot_2018}. With the so obtained configuration, for each pair of arms we project their coordinates on the $xy$ plane and count how many times they cross each other. 
Once the linking numbers between a certain arm and all the others are calculated, we take their absolute values and sum them, in order to have a measure of how strongly each arm is linked.\\
This procedure works well in the case of closed arms, while in the case of doubly grafted arms we have to take some additional precaution, as the linking number is only defined for closed curves. There are many ways to close a curve and this closure can determine the linking number, so this is a crucial step. For each pair of arms, we decide to take the average linking number between two different closures, where we alternatively close one arm by connecting its anchoring points with an arc laid on the surface of the core and close the other arm directly through the inside of the core (see Fig.\ref{fig:Lk_sketch}b).  The linking number so defined is symmetrical with respect to the two arms, but not necessarily an integer number.
\section{\label{sec:results}Results}
\subsection{\label{sec:unlinked}Linear star polymers and unlinked cyclical star polymers.}
We consider, in this section, polymer stars where the grafted chains are either linear or unlinked, unknotted rings. For the sake of simplicity, we refer to them as linear stars and cyclical stars, respectively. 
We aim to characterize the physical properties of a cyclical star with respect to those of a linear one, for which computational and theoretical results already exist. We focus now on the {\it gyration radius} $R_g$ (see Sec.~\ref{ssec:asph}) and the {\it radial density} $\phi(r)$, defined as:
\begin{eqnarray}
\label{phi_r}
    \phi(r) &\equiv& \frac{1}{4\pi r^2}\left\langle\sum_{i=1}^{N-1}\delta\left( r-|\textbf{r}_i-\textbf{r}_0|\right)\right\rangle,
\end{eqnarray}
where $N=f\cdot n +1$  is the total number of beads, $\textbf{r}_0$ is the position of the core and $\langle\cdot\rangle$ is the ensemble average, computed over time along a single, long trajectory (see Sec.~\ref{sec:methods}). Furthermore, we investigate the degree of symmetry of the polymers, by measuring their asphericity $A$ for which, to the best of our knowledge, no theoretical or numerical results are present. For the definition of $A$, See Methods, \ref{ssec:asph}.
\subsubsection{\label{ssec:gyration}Metric properties.}
The scaling behavior for the gyration radius of a linear star polymer was theoretically computed by Daoud and Cotton in Ref.~\cite{daoud_star_1982}. As we focus on self-avoiding stars in good solvent, with relatively few and long arms (i.e. we always consider $f^{1/2}\leq n$), the macromolecules are always in the so-called \textit{swollen} regime. In practice, this means that that the arms always remain in semi-dilute conditions and excluded volume interactions are present~\cite{grest_structure_1987}. In this regime, given a Flory exponent of $\nu=0.588$, $R_g$ behaves according to: %
\begin{equation}
    R_g \propto  n^{\nu}f^{(1-\nu)/2}= n^{.588}f^{.206}.
\end{equation}
%
\begin{figure}
\includegraphics[width=0.48\textwidth]{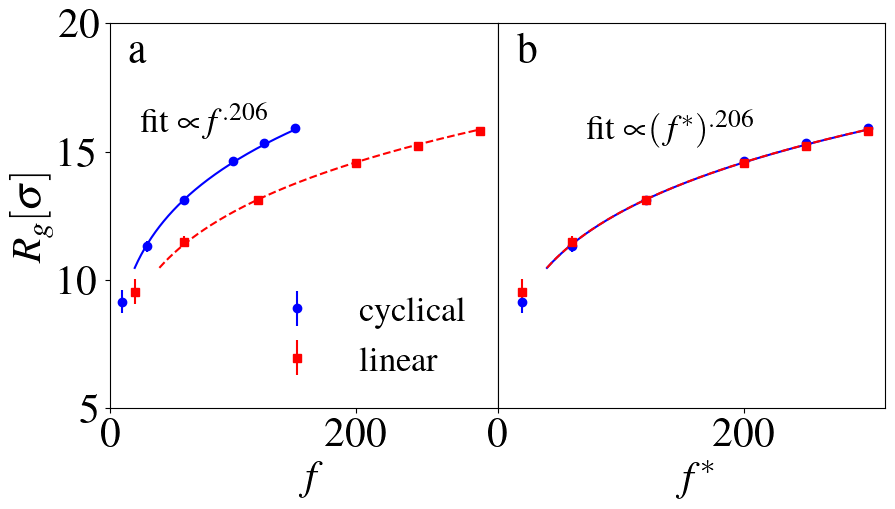} 
\caption{\label{fig:Rg_n100} Gyration radius $R_g$ for stars with linear (red squares) and cyclical (blue circles) arms with $n_L$ = 50 and $n_C$=100, respectively a) as a function of $f$ and  b) as a function of $f^*$, where $f^*=2f_C$ for cyclical stars and $f^*=f_L$ for linear stars. In both panels, lines are fit to the power law $f^{.206}$, scaling typical  of linear star polymers in the swollen regime.}
\end{figure}

In the case of linear stars, we consider $20\leq f_L \leq 300$ arms of $n_L=50$ beads per arm. Conversely, for cyclical stars  we set $10\leq f_C \leq 150$ arms and $n_C=100$ monomers per arm. We compare the gyration radius for the systems mentioned above: within the chosen range of values of the parameters we observe a correspondence between the gyration radius of a cyclical star polymer having $f_C$ arms and $n_C$ monomers per arm and a linear star polymer with $f_L = 2 f_C$ and $n_L = n_C/2$. Indeed, looking at Fig.~\ref{fig:Rg_n100}a, we immediately notice how the linear star polymers follow the scaling law $R_g\simeq f^{.206}$ and, interestingly, how unlinked cyclical star polymers feature the same behavior. Upon normalizing their functionality as $f^* = 2 f_C$ (i.e. we consider, for each cyclical arm, a functionality of 2 instead of 1) and $f^*=f_L$, the two sets of data collapse (see Fig.~\ref{fig:Rg_n100}b), with a slight disagreement at very small values of $f^*$, due to finite size effects.
\begin{figure}
\includegraphics[width=0.48\textwidth]{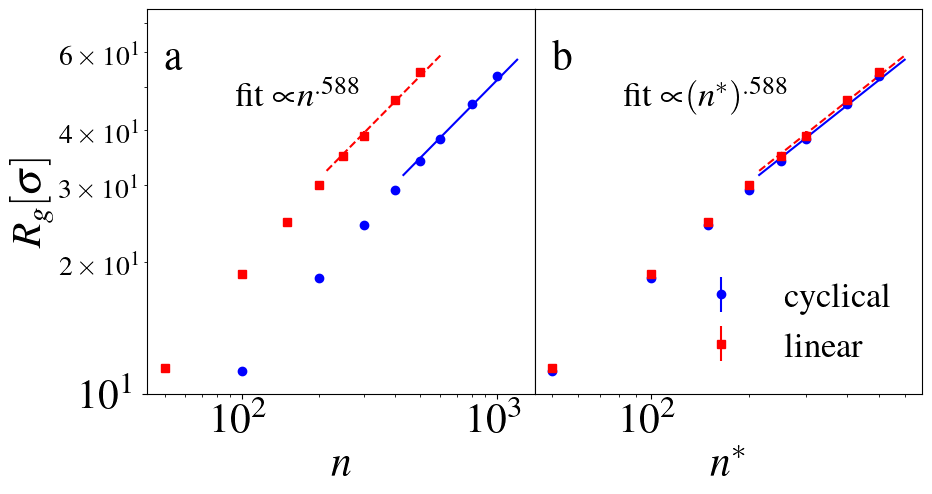}
\caption{\label{fig:rg_n} Gyration radius $R_g$ for star polymers with linear (red squares) and cyclical (blue circles) arms with $f_L$ = 120 and $f_C$=60, respectively a) as a function of $n$ and  b) as a function of $n^*$, where $n^*=n_C/2$ and  $n^*=n_L$. In both panels, lines are fit to the power law $(n^*)^{.588}$.} 
\end{figure}

We recover the features of the swollen star regime also upon varying $n$ at fixed $f^*=60$, as shown in Fig.~\ref{fig:rg_n}: we observe that data for different values of $n^*$, where $n^*=n_C/2$ for cyclical and $n^*=n_L$ for linear arms, collapse on the same line; the data are proportional to $(n^*)^{0.588}$ for sufficiently large values of $n^*$. These results show that the equivalence between unlinked rings and linear chains, valid for flat cyclical brushes~\cite{Reith_2011,Jehser2019}, holds also for a star architecture. 
\begin{figure}
\includegraphics[width=0.48\textwidth]{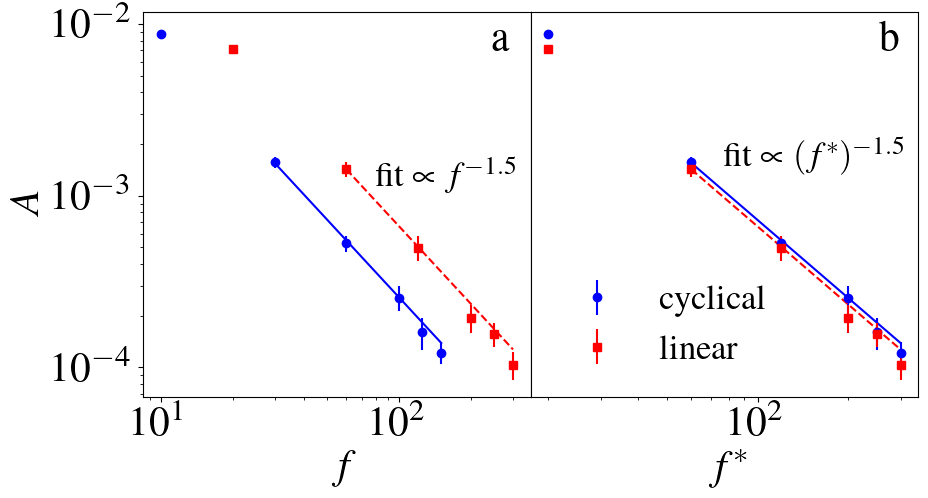}
\caption{\label{fig:asph_n} Asphericity $A$ for star polymers with linear (red squares) and cyclical (blue circles) arms with $n^*$ = 50 as a function of $f$ (a) and $f^*$ (b). $A$ shows a decay proportional to $(f^*)^{-1.5}$ for both cyclical and linear stars }
\end{figure}

Finally, we analyze the shape of the polymer, by looking at the asphericity.
Since we consider stars having all arms of the same length, we can expect that increasing the number of arms will result in a more spherically symmetric macromolecule. We confirm this by plotting the asphericity $A$ as a function of $f$ in Fig.~\ref{fig:asph_n}a: as $f$ grows, $A$ decays with the power law $f^{-1.5}$. Similarly to the gyration radius we notice that, when normalizing to $f^*$, the data points for cyclical and linear stars fall on the same master curve, especially for large values of $f^*$ (Fig.~\ref{fig:asph_n}b).
\subsubsection{\label{ssec:radial}Radial density.}
Another theoretical result of Ref.\cite{daoud_star_1982} is the radial density profile $\phi(r)$ of a star polymer. In the swollen regime,  $\phi(r)$ goes as \cite{grest_structure_1987}:
\begin{equation}
\label{eq:radprof}
    \phi(r) \propto f^{(3\nu-1)/2\nu}r^{(1-3\nu)/\nu}=f^{.650}r^{-.130}
\end{equation}
for $(\sigma_c+\sigma)/2 < r < r_{cor}$, where $r_{cor}$ is the so-called corona radius, defined as the radius of the sphere centred in the core that   contains all the monomers:
\begin{equation}
  4 \pi \int_0^{r_{cor}} r^2 \phi(r) dr = Nf.  
\end{equation}
In practice, the corona radius sets the value of $r$ at which the finite size of the macromolecule becomes apparent: the power law $r^{(1-3\nu)/\nu}$ becomes an exponential decay. For linear stars in the scaling regime, one can show that in the limit of a small core $r_{cor}\simeq\left(5/11\right)^{1/2}R_g$ \cite{grest1996star,likos2001effective}, i.e. $r_{cor}$ and $R_g$ are proportional.

\begin{figure}
\includegraphics[width=0.44\textwidth]{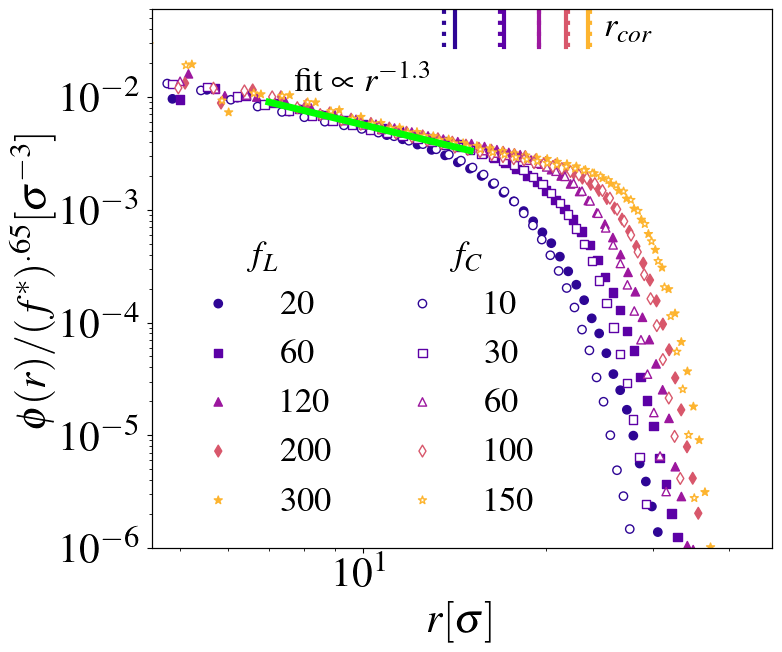}
\caption{\label{fig:rad_n100} Comparison of   radial density $\phi(r)$ for linear and cyclical star polymers with $n=50$ beads for the linear stars, $n=100$ beads for the cyclical one and various values of the normalized functionality $f^*=f_L=2f_C$. The curves are normalized with $(f^*)^{-.65}$ and fitted to the power law $r^{-1.3}$, typical of the density profile of linear star polymers in the swollen regime. The corona radius $r_{cor}$ is shown at the top of the figure, with a solid line for linear stars and a dotted one for cyclical stars. The $\phi(r)$ of both cyclical and linear stars has the same behavior except for values of $r$ larger than $r_{cor}$, where the decay of the $\phi(r)$ of cyclical star is steeper than its linear counterpart.}
\end{figure}

The swollen regime is present for both linear and cyclical stars, as can be seen in Fig.~\ref{fig:rad_n100}. Upon considering the normalized functionality $f^*$ and arm length $n^*$, we notice that the difference in radial density is minimal between linear and cyclical stars; at fixed $f^*$ the data sets show a discrepancy only for very large values of $r$, after the onset of the final decay.
More specifically, the $\phi(r)$ of cyclical stars decays faster, as the two halves of the ring arms must reconnect and, as such, they are less likely to extend as far as the linear counterparts.  All in all, the differences  becomes less distinct upon increasing $f^*$, which further corroborates the mapping of $f^*$ and $n^*$ proposed above. After normalization also the corona radius shows good agreement between linear and cyclical stars, with the cyclical $r_{cor}$ being slightly smaller than the linear one for a small $f^*$ and slightly larger for a large $f^*$. We highlight that equally small differences can be observed in Fig.~\ref{fig:Rg_n100}b and suggest the same proportionality between $r_{cor}$ and $R_g$, valid for linear stars, holds also for cyclical stars. This can be rationalized by the cyclical constraint: the need for reconnecting cyclical strands is reducing $r_{cor}$ at small $f^*$ while, as they are grafted to the core by a single monomer, the two halves of ring arms can stretch slightly further than linear ones when $f$ (and, consequently, the density close to the core) becomes very large. 
%
%
%
\subsection{\label{sec:linked}Linked cyclical star polymers.}
The emergence of links between the chains depends on the route of chemical synthesis employed to create the macromolecule. If the star is assembled by grafting adsorbed rings on the surface of the core, then no permanent link can be formed. However, this strategy may not be always suitable, depending on the substrate and on the particular application. Another possible route may involve the closure of linear arms, for example through click chemistry~\cite{binder2007click,geng2021click}. In that case one can imagine that links between the arms may form, especially in the limit $n \gg 1$. 

Here we do not explore a specific assembly mechanism, but rather tackle the problem of characterizing the properties of a cyclical stars with concatenated arms. As already mentioned, we consider two different protocols: i) temporarily turn off the self-avoidance among arms and confine the system; ii) consider ``doubly grafted'' linear chains and, by employing a  geometrical rule, construct the chains with a certain degree of linking.

The introduction of links between the cyclical arms induces permanent topological constraints and \textit{fixes} the topology of the star. This makes the computational study more challenging as one must not average simply over time but also
over different realisation of the topology (see Methods section, Sec.~\ref{sec:methods}). Indeed, the quantity we use to characterise the topology of the macromolecules 
is the {\it linking number} $Lk$, defined in Methods, Sec.~\ref{ssec:linkn}, which gives a measure of how many times two closed curves are intertwined. Independently from the route we follow to introduce the links, the topology will be fixed at the steady state; as such, the linking number will be constant and, in order to collect statistics, we have to produce different realisations of the linked stars. 
\subsubsection{\label{ssec:compression}Linking by confinement of phantom rings.}
\begin{figure}[!t]
\includegraphics[width=0.48\textwidth]{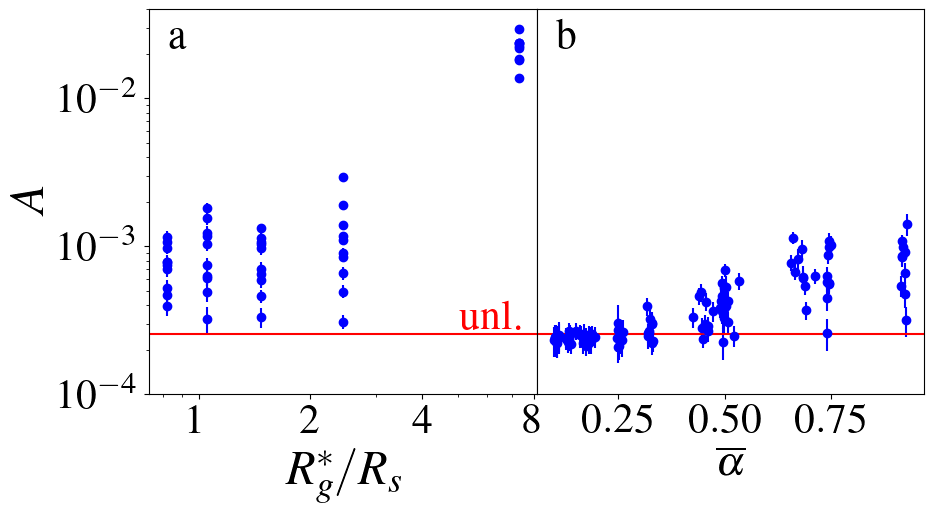}
\caption{\label{fig:asph_lk} Asphericity $A$ as a function of  a)  the inverse confinement radius $R_s^{-1}$ (for compressed stars) and b) the mean opening angle $\overline{\alpha}$ (for stars with doubly grafted arms). These data are taken for cyclical stars with $n_C=100$ and $f_C=100$, and compared to the asphericity of the corresponding unlinked cyclical star (red line).}
\end{figure}

The first protocol we employ to obtain stars of linked rings takes inspiration from equilibrium sampling techniques of knots ~\cite{tubiana2013spontaneous}, as well as strategies to model the action of topoisomerares in the genome~\cite{michieletto_is_2015,michieletto2022dynamic}.
We describe the procedure in detail in Sec.~\ref{sec:methods}): briefly, we remove the steric repulsion among non-neighboring monomers, set a spherical confinement and wait for the system to reach a steady state.  We then reintroduce the self-avoidance gradually: once the latter is fully restored, we obtain a system with a fixed topology.
In order to tune the amount of concatenation, we confine the system employing a steric spherical confinement, as mentioned (see Methods, Sec.~\ref{ssec:compr}). We take inspiration here from the modelisation of microgels, where spherical confinement is effectively able to control the internal structure of similarly cross-linked network~\cite{gnan2017insilico}.\\
\begin{figure}[!t]
\includegraphics[width=0.48\textwidth]{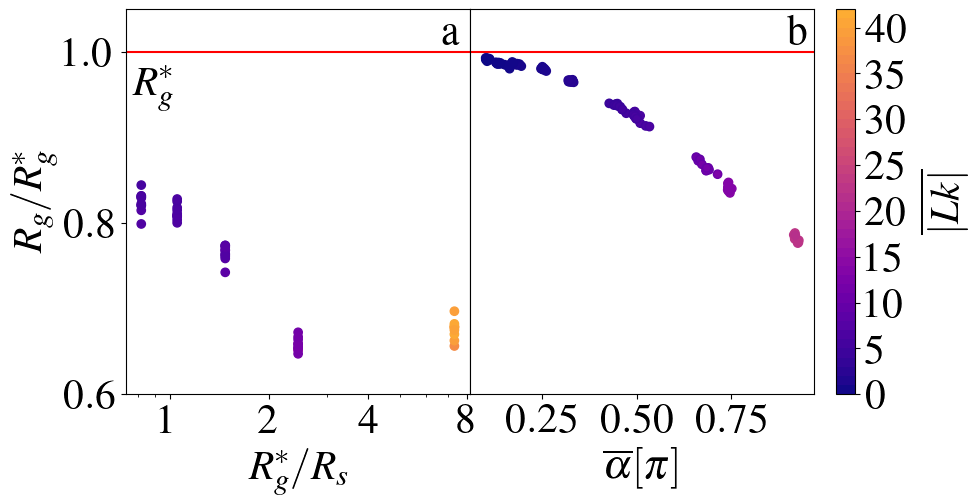}
\caption{\label{fig:lks_col} Gyration radius $R_g$ as a function of  a)  the inverse confinement radius $R_s^{-1}$ (for compressed stars) and b) the mean opening angle $\overline{\alpha}$ (for stars with doubly grafted arms), scaled by the gyration radius of an unlinked cyclical star polymer $R_g^*$. The colors show the average absolute linking number per arm $\overline{\vert Lk\vert}$. These data are taken for cyclical stars with $n_C=100$ and $f_C=100$. 
}
\end{figure}
We report in Fig.~\ref{fig:lks_col}a the gyration radius as a function of the radius of the spherical confinement $R_s$. Each point reported in the plot represents an  independent, random realisation of a  topology; the color code represents the average absolute linking number $\overline{\vert Lk\vert}$ per arm, where $\overline{(\cdot)}$ denotes the average over all the arms of the star. Here we fix $f_C=100$ and $n_C=100$; notice that both $R_g$ and $R_s$ are normalized by $R_g^*$, the average gyration radius of an unlinked cyclical star with $f_C$ arms of $n_C$ monomers per arm. We see that, even in absence of compression ($R_g^*/R_s\ll 1$), allowing the arms to freely cross each other leads to a significant decrease of $R_g$ ($\simeq 80\%$ of the unlinked cyclical gyration radius $R_g^*$), as the absolute linking number becomes immediately larger than zero ($\overline{\vert Lk\vert}\simeq 10$). Upon forcing the system to fit a spherical confinement comparable or smaller than its (unlinked) average size, we observe an increase of $\overline{\vert Lk\vert}$, leading to a strong decrease of the gyration radius. When a sufficiently strong confinement ($R_g^*/R_s>2$) is imposed, upon restoring the self-avoidance, the cyclical star becomes more visibly aspherical; indeed the asphericity (see Fig.\ref{fig:asph_lk}a) grows by one order of magnitude with respect to the unlinked case, at fixed $f_C$. In addition, its gyration radius becomes independent of $R_g^*/R_s$, assuming constant value $R_g/R_g^* \approx 0.6$. Notably, the emergence of asphericity is purely a consequence of the topological complexity, since the application of the radial confinement does not violate such symmetry. In agreement with this observation, at the largest value of $R_g^*/R_s$ considered, we observe a huge value of the average linking, at $\overline{\vert Lk\vert} \simeq 40$.  
\subsubsection{\label{ssec:anchoring}Linking with doubly grafted arms.}
An alternative approach for introducing topological entanglements in cyclical stars is to mimic the result of a cyclization process between pairs of arms. In our approach, we consider arms whose extremities are both anchored at different position on the surface of the core. As already mentioned, this kind of ``doubly grafted'' cyclical arm can be imagined as the result of e.g. a click chemistry approach by which the free extremities of two separate linear arms form a bond. Alternatively, the same chemical reaction can take place between the free extremity of a linear arm and the core. 

In our approach, we encode the topology of these polymers in the initial configurations, creating the doubly-grafted arms via a geometrical construction (see Methods, Sec.~\ref{ssec:init}). We then classify the results as function of a geometrical parameter, the average opening angle $\overline{\alpha}$, that is the average angle between the two anchoring points of the arms with respect to the center of the core. We consider again star polymers with $f_C=100$ and $n_C=100$ and plot their gyration radius, normalized again by $R_g^*$, as a function of $\overline{\alpha}$ in  Fig.~\ref{fig:lks_col}b). As before, each point reported represents an independent, random realisation of a topology; the color represents the average absolute linking number $\overline{\vert Lk\vert}$ per arm. 
\begin{figure}
\includegraphics[width=0.44\textwidth]{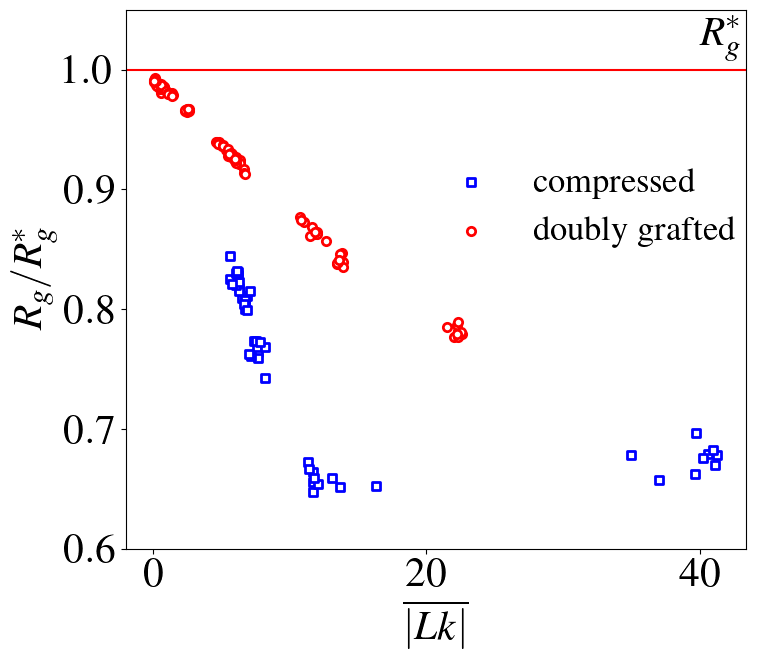}
\caption{\label{fig:lks_rgs} Gyration radius $R_g$ of linked cyclical stars with $f_C=100$ and $n_C=100$ as a function of the average absolute linking number per arm $\overline{\vert Lk\vert}$, scaled by the gyration radius of an unlinked cyclical star polymer $R_g^*$. 
While both kinds of stars display a reduction in $R_g$ as $\overline{\vert Lk\vert}$ grows, their behaviors are very different.
}
\end{figure}
We notice that, upon increasing $\overline{\alpha}$, stars become more compact with respect to the unlinked reference value; further, we observe that controlling $\overline{\alpha}$ allows to control $\overline{\vert Lk\vert}$. So, in agreement with the previous method, an increase of the absolute linking number correlates with a decrease of the gyration radius. However, for very small values of $\overline{\alpha}$ we basically recover the unlinked case, with $R_g/R_g^* \simeq 1$ and a vanishing $\overline{\vert Lk\vert}$. Also, the overall reduction of the star's size 
is not as large as in the confined case ( Fig.~\ref{fig:lks_col}a)): for the largest value of $\overline{\alpha}$ , $R_g \simeq 80\% R_g^*$. Furthermore, while the asphericity of the star increases on average with $\overline{\alpha}$ (see Fig.~\ref{fig:asph_lk}b), it always remains more than an order of magnitude smaller than that of highly compressed stars (see Fig.~\ref{fig:asph_lk}a) and comparable with the value obtained for unlinked star of comparable size (see Fig.~\ref{fig:asph_n}). 
Indeed, comparing $R_g/R_g^*$ as a function of the linking number for both methods 
as shown in Fig.~\ref{fig:lks_rgs}, reveals two completely different behaviors. While the gyration radius consistently remains a decreasing function of the absolute linking number, in the case of doubly grafted arms this correlation is essentially linear and relatively mild. On the contrary, in the confined case, $R_g/R_g^*$ initially decreases also linearly, albeit strikingly faster, and then reaches a plateau around $\overline{\vert Lk\vert} \approx 15$.\\
\subsubsection{\label{ssec:lkscaling}Scaling for linked cyclical stars.}
\begin{figure}
\includegraphics[width=0.44\textwidth]{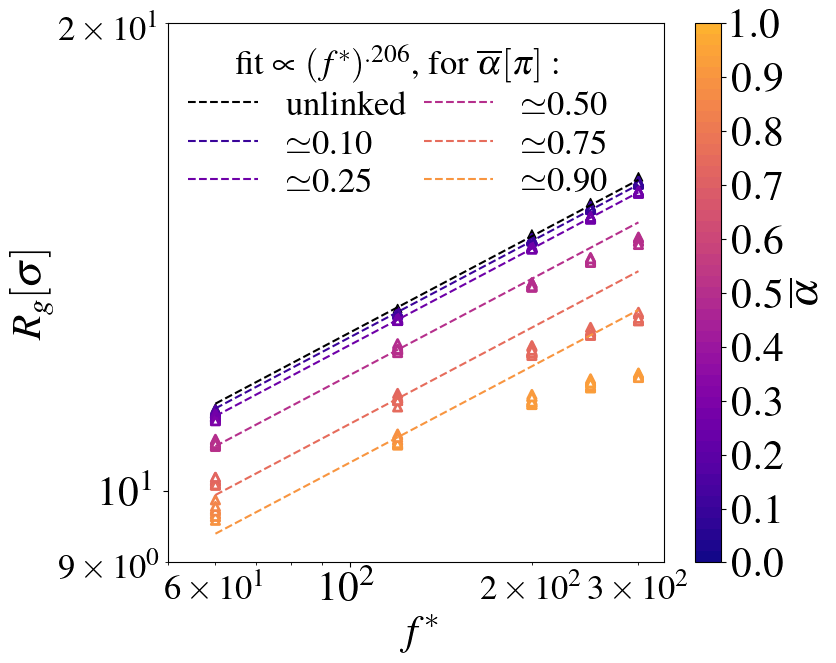}
\caption{\label{fig:rg_seed} Gyration radius $R_g$ of linked cyclical stars with $n_C=100$ as a function of $f^*$ for different values of the mean opening angle $\overline{\alpha}$. As the $(f^*)^{.206}$ regime breaks down for large values of  $\overline{\alpha}$ and $f^*$, the fits shown have been calculated only for small values of $f^*$  ($f^*<=120$). 
}
\end{figure}
In this last section, we will show how the presence of topological entanglements has an effect also on the scaling properties of the gyration radius. We will focus, mostly, on the ``doubly grafted'' arms model, as it is, overall, the simplest of the two methods and possesses the best connection to experiments.\\
We report the gyration radius of doubly grafted stars as a function of the functionality $f^*=2f_C$ in Fig.~\ref{fig:rg_seed}, for different values of the opening angle $\overline{\alpha}$.
We observe that imposing a large $\overline{\alpha}$ that corresponds, as seen previously, to a large $\overline{\vert Lk\vert}$, causes the $(f^*)^{.206}$ scaling law to break down. In particular, $R_g$ becomes almost insensitive of $f^*$ in strongly linked stars.\\ 
\begin{figure}
\includegraphics[width=0.44\textwidth]{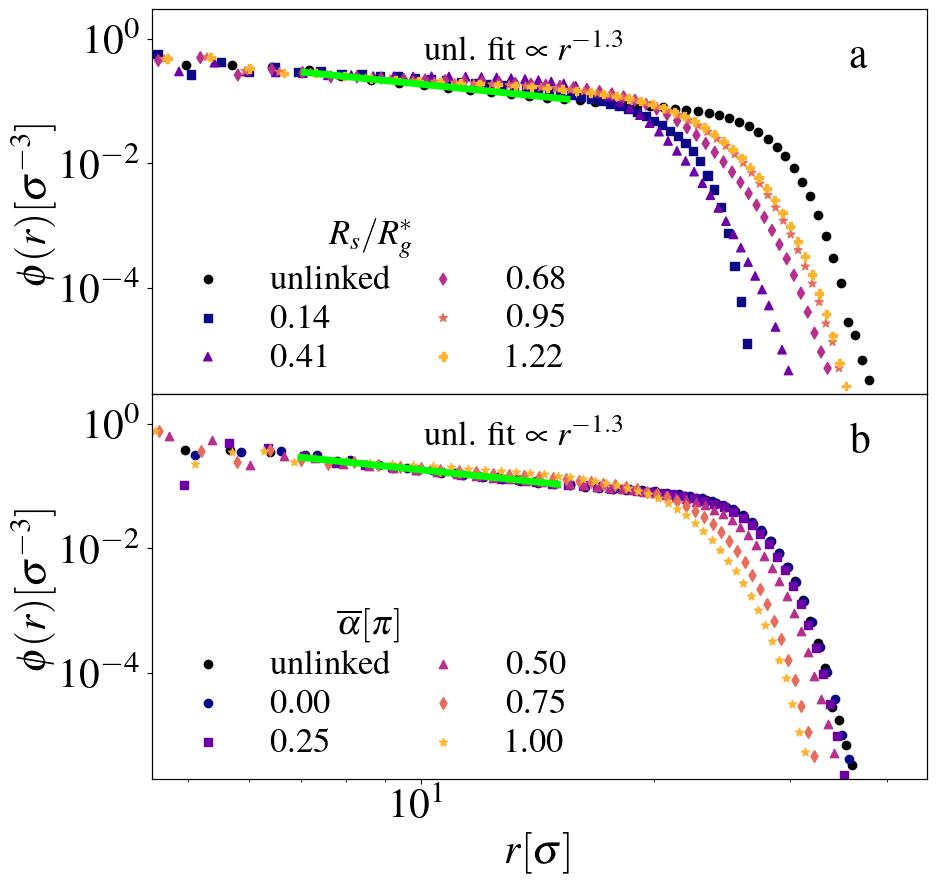}
\caption{\label{fig:rad_seed} Radial density $\phi(r)$ of linked cyclical stars with $f=100$ and $n=100$ for different values of a)  the confinement radius $R_s$ (for compressed stars) and b) the mean opening angle $\overline{\alpha}$ (for stars with doubly grafted arms). The radial density of a cyclical star of unlinked rings and its $\propto r^{-1.3}$ fit is shown for reference in both figures.}
\end{figure}
Concurrently, we observe that the radial density profile of linked cyclical stars does not follow the behaviour described by Eq.~\ref{eq:radprof} at high values of $\overline{\vert Lk\vert}$. Specifically, the radial density profile become almost flat at intermediate values of $r$ and decays more sharply at the level of the external corona, signalling an overall more compact conformation. This change in regime is gradual for stars with doubly grafted arms, with the profile slowly departing from $r^{-1.3}$ as $\overline{\alpha}$ grows (see Fig.\ref{fig:rad_seed}a), while it is more sudden in the case of compressed stars (Fig.\ref{fig:rad_seed}b).\\
The significant discrepancies between the $R_g$ and $\phi(r)$ behaviors of doubly-grafted and compressed stars show how different linking methods can define very different star configurations. In fact, in contrast to the dynamical approach of phantom rings, the geometrical linking approach of doubly-grafted stars does not allow for an absolute linking number larger than 1 between two arms, as each arm initially delimits a convex 2D polygon. Furthermore, phantom rings can not only form higher order links with other arms, but also tie themselves into knots, yielding much more crumpled configurations.
\section{\label{sec:outro}Discussion and conclusions.}
In this work we have reported on the metric and topological properties of cyclical star polymers. We show that unlinked cyclical stars possess the same scaling properties than linear stars; we further show that their internal structure is effectively the same, with differences only appearing at large distances from the core, where the cyclical nature plays a more important role. The comparison can be carried out only if the cyclical nature is properly taken into account, by rescaling the number of arms $f$ and the number of monomer per arms $n$. Essentially, this amounts to say that each cyclical arm counts as two half-length linear arms:
\begin{equation}
\label{eq:fn}
    f_R = 2 f_L \qquad n_R = 1/2 n_L.
\end{equation}
The overall shape of the macromolecule, assessed via the asphericity, obeys the same scaling in both systems; remarkably, we report a scaling law for the asphericity itself that, to the best of our knowledge, has been never highlighted.\\
The simple relationship between linear and cyclical arms (Eq.~\ref{eq:fn}) has been observed first in plane cyclical brushes and remains valid also here. This means that the blob picture holds in both cases even if the arms are cyclical, as long they are unlinked.\\
When the topology becomes non-trivial, via the introduction of permanent links, the metric properties of the macromolecules change drastically. We consider two different strategies to introduce links in an effective fashion, either removing the self avoidance under confinement and re-introducing it after a new equilibrium is reached or constructing ``doubly grafted'' arms. The first strategy may be regarded as a bit extreme and is, possibly, not very realistic from the perspective of chemical synthesis. Further, it allows for the formation of multiple links and knots, yielding very crumpled configurations. However, it is interesting to consider it as a ``worst case'' scenario, that may be approached if the synthesis is made in bad solvent conditions. Conversely, the second strategy aims to represent the results of an assembly in good solvent achieved via a click chemistry approach. It yields considerably simpler complex topologies and the effects are definitely milder. It is nonetheless interesting to notice that, qualitatively, the results are the same, i.e. the introduction of links changes both the metric and the scaling properties of the star in a considerable fashion, yielding an overall crumpled and more compact macromolecule.\\Our findings open up different questions for future research. The properties of (unlinked) cyclic planar brushes, such as enhanced steric stabilization, lubrication and nonspecific adsorption, have recently emerged both in experiments and in simulations\cite{morgese2018cyclic,roma_theoretical_2021,galata_topological_2024}; they are a consequence of the non-trivial interaction properties stemming from the cyclical nature of the arms. Since, as mentioned in the introduction, the effective interaction governs the phase diagram of the system, it would be interesting to investigate the properties of the effective interaction between cyclical stars. For example, from an applicative perspective, this could lead to improvements in the performance of lubricants, as possibly suggested by the planar brush case. A very different application may be as a ``cleaning'' agent, to absorb and remove pollutants from water; linear star polymers have been already considered in this context\cite{roma2021theoretical}. One interesting aspect here is that adsoption has been shown to modify the properties of the polymer both in theory and in experiments, for different architectures\cite{roma2021theoretical,EskandariNasrabad2023,vorsmann2024colloidal}. In all cases, a transition to a bad solvent conformation seems to take place. It would be interesting to check if the cyclization of the arms or the introduction of a more complex topology could affect this phenomenon, allowing, for example, for a higher absorption capability or, conversely, for an improved selectivity.

\begin{acknowledgments}
This work has been supported by the project "SCOPE - Selective Capture Of metals by Polymeric spongEs" funded by the MIUR Progetti di Ricerca di Rilevante Interesse Nazionale (PRIN) Bando 2022 - grant 2022RYP9YT. 
All authors acknowledge the CINECA award under the ISCRA initiative, for the availability of high performance computing resources and support. 
D.B. and L.T. acknowledge the computing resources offered by the UniTN HPC cluster.
\end{acknowledgments}

\bibliographystyle{apsrev4-2}
\bibliography{cyclical}

\end{document}